\documentclass{svproc}
\usepackage{comment}
\usepackage{multirow}  
\usepackage{xcolor}
\usepackage{soul}
\usepackage{amsmath}
\usepackage{algorithm}
\usepackage{algpseudocode}
\usepackage{graphicx}
\usepackage{siunitx}
\usepackage[table]{xcolor}
\usepackage{adjustbox}
\usepackage{adjustbox}
\usepackage{colortbl}
\usepackage{subfig}
\usepackage{siunitx} 
 
\usepackage[T1]{fontenc}
\usepackage[hyphens]{url}
\usepackage{hyperref}
\usepackage[numbers]{natbib} 
\hypersetup{breaklinks=true}
\usepackage{amsmath,amssymb,amsfonts}
\usepackage{float}
\usepackage{booktabs} 

\begin{document}
\mainmatter              
\title{The Role of Community Detection Methods in Performance Variations of Graph Mining Tasks}
\titlerunning{Impact of Community Detection on Task Performance}  
%
\author{Shrabani Ghosh\inst{[0000-0002-6084-4964]} \and Erik Saule\inst{[0000-0003-1634-9234]}}
%
\authorrunning{Ghosh et al.} 
%
%
\institute{University of North Carolina Charlotte, Charlotte NC 28223, USA.\\
\email{\{sghosh15,esaule\}@charlotte.edu}}

\maketitle              

\begin{abstract}

In real-world scenarios, large graphs represent relationships among entities in complex systems. Mining these large graphs—often containing millions of nodes and edges—helps uncover structural patterns and meaningful insights. Dividing a large graph into smaller subgraphs facilitates complex system analysis by revealing local information. Community detection extracts clusters or communities of graphs based on statistical methods and machine learning models using various optimization techniques. Structure-based community detection methods are more suitable for applying on graphs because they do not rely heavily on rich node or edge attribute information. The features derived from these communities can improve downstream graph mining tasks, such as link prediction and node classification. In real-world applications, we often lack ground-truth community information. Additionally, there is neither a universally accepted “gold standard” for community detection nor a single method that is consistently optimal across diverse applications. In many cases, it is unclear how practitioners select community detection methods, and choices are often made without explicitly considering their potential impact on downstream tasks. In this study, we investigate whether the choice of community detection algorithm significantly influences the performance of downstream applications. We propose a framework capable of integrating various community detection methods to systematically evaluate their effects on downstream task outcomes. Our comparative analysis reveals that specific community detection algorithms yield superior results in certain applications, highlighting that method selection substantially affects performance. 

\end{abstract}
\section{Introduction} \label{intro}

Graphs, or networks, are fundamental structures used to model numerous complex systems across a wide range of real-world contexts, from social media networks to biological networks. Graphs represent relationships between entities, with nodes signifying individual entities and edges connecting those related entities. For graph mining, feature extraction is crucial for downstream tasks, which can often be challenging. To address this issue, a large graph is often partitioned into smaller subgraphs, facilitating easier processing and mining. Community detection~\cite{su2022comprehensive} aims to uncover groups of nodes that are more densely connected within the same community than the rest of the graph. Researchers use community detection for downstream tasks such as node classification and link prediction~\cite{han2023link}. In real-world applications, labeled datasets are often rare; by employing graph features, nodes/edges can be automatically classified using community features~\cite{bhagat2011node}. In node classification, unlabeled nodes are classified using community features. In link prediction, predicting relationships that do not exist but could be created in the future can be assessed using community features~\cite{zhang2018link}. 

Several approaches to community detection have been proposed, and surveys have provided a summary and comparative analysis of these methods~\cite{rehman2012graph,su2022comprehensive,jin2021survey}. The wide variety of community detection algorithms available based on various approaches, such as modularity optimization~\cite{von2007tutorial}, spectral~\cite{von2007tutorial}, and graph embedding techniques~\cite{perozzi2014deepwalk,cui2020adaptive} to extract features from graphs as communities and utilize the features for further graph analysis. Researchers often rely on choosing methods without evaluating how their choice may affect the performance of downstream applications. This oversight can lead to suboptimal results and missed opportunities to maximize effectiveness in real-world scenarios. To evaluate the impact of community detection algorithms on downstream applications, we selected several well-known methods that provides comprehensive coverage of non-overlapping and overlapping approaches. For downstream analysis, we chose representative applications that heavily rely on community structures at the intermediate level of their processes: 1)~Recommendation Systems that leverage community propensity information to generate personalized product recommendations based on group interests; 2)~Trust Prediction which utilize community structures in social graphs to predict absent or potential link/trust relationships among users; and 3)~Anomaly Detection to identify irregularities or abnormal interaction patterns within networks to classify potentially malicious nodes.

Our contribution is to systematically explore the relationship between the choice of community detection algorithms and the performance outcomes in applications. We make the following contributions: 1)~Investigating whether the choice of community detection methods has an impact on downstream task performance. 2)~We introduce a framework capable of integrating and evaluating different community detection algorithms, thus facilitating direct comparisons in downstream applications. 3)~We experimentally validate that the selection of community detection algorithms significantly impacts downstream task performance, highlighting that certain methods are inherently more suitable for particular applications and datasets. 

In this paper, Section~\ref{sec:CD} discusses community detection methods briefly. Section~\ref{sec:framework} describes our framework for study the impact of community detection method in downstream tasks. We then study our three use cases: Recommendation (Section~\ref{sec:usecase:recommendation}), Trust Prediction (Section~\ref{sec:usecase:trust}) and Anomaly Detection (Section~\ref{sec:usecase:anomaly}). Section~\ref{sec:ccl} presents the conclusion.

\section{Community Detection}
\label{sec:CD}
\subsection{Definitions}
A basic network of $n$ nodes and $m$ edges is defined as $G = (V,E)$, where $V = \{v_1,v_2,...,v_n\}$ is the node set and $E=\{e_{ij}\}$ represents the edge set, when there is an edge exists between node $v_i$ and $v_j$. $A=[a_{ij}]$ denotes an $n \times n$ dimensional adjacency matrix where $a_{ij} = 1$, if $e_{ij}$ belongs to $E$, otherwise, $a_{ij} = 0$. A degree of a node $v$ denotes by $deg(v)$. 

Community detection, sometimes termed partitioning or clustering, is referred to as a community detection method in this paper. 

\subsubsection{Communities:} A community in a graph is a concept intended to denote a subset of nodes that are more densely connected to each other than to the rest of the network. These nodes form tightly-knit groups, where intra-group connections are significantly stronger or more frequent compared to inter-group connections. The nodes in a community often have a regional structure and some common group properties. We define number of communities as $k$. The set of communities is denoted by $C = \{C_1,C_2,C_3, ..., C_k\}$. Communities can be either non-overlapping or overlapping. In non-overlapping communities, nodes belong to only one cluster, whereas in overlapping communities, nodes can belong to multiple communities. The community (or communities) node $i$ belongs to is denoted $C(i)$.

\subsubsection{Quality Functions}
The community detection algorithms identifies the subgraphs using objective functions. Some algorithms return a few communities, while others return many communities. The detected communities are typically assessed using quality functions, which consider two perspectives: 

\paragraph{Global Partition Quality} measures how well the entire graph is partitioned. A common example is modularity, which quantifies the density of edges inside communities compared to edges between communities.
\begin{equation}
   Modularity = \frac{1}{2m}\sum_{(u,v) \in V^2} \left[A_{uv} - \left(\frac{deg(u)deg(v)}{2m}\right)X_{uv}\right]
\end{equation}

The decision variable is defined by $X_{uv} \in \{0,1\}$ which indicates whether the nodes $u$ and $v$ belong to the same cluster ($X_{uv}$ = 1) or not ($X_{uv}$ = 0). The modularity score ranges within -1 to 1. A value close to 1 denotes a strong community structure.  
\paragraph{Local Community Quality} measures how cohesive individual communities are, often using metrics such as density or conductance. Density measures the proportion of actual edges present within a community compared to the maximum possible number of edges. The density of a community with $n$ nodes and $m$ internal edges given by:
\begin{equation}
    Density = \frac{2m}{n(n-1)}
\end{equation}  
The density value ranges between 0 and 1. 
There are some other quality functions: the normalized cut that measures the cost of cutting edges between communities while keeping internal connections within communities~\cite{dhillon2004kernel}, Clustering Coefficient assesses the triangular pattern and the connectivity in a vertex neighborhood: a vertex has a high Clustering Coefficient~\cite{saramaki2007generalizations} if its neighbors tend to be directly connected to each other. More quality functions can be found in~\cite{chakraborty2017metrics}

\subsection{Community Detection Algorithms}
Many  algorithms have been  designed to identify communities. They fall within various categories from hierarchical clustering to overlapping, non-overlapping, or even machine learning models. We selected five methods that are purely based on the topology of the graph, allowing us to apply them to any application. 
We briefly explain the methods below. 

\paragraph{Louvain}\cite{blondel2008fast}: The Louvain method uses modularity as an objective function to generate communities through an iterative process. Initially, it assigns each node to a different community and then greedily moves the nodes from one community to another by optimizing the modularity.

\paragraph{Spectral Clustering}\cite{donath1973lower}: Spectral methods use particular top eigenvectors of a matrix derived from the distance between points. This method begins by constructing an affinity matrix that reflects the similarity between points using a Gaussian kernel. The diagonal elements of the matrix are set to zero, ensuring that self-similarity is not considered. From the affinity matrix, a graph Laplacian is computed, and its top eigenvectors are used to embed the nodes into a lower-dimensional space. Finally, a standard geometric clustering algorithm, such as $k$-means, is applied to the embedding to partition the points~\cite{schenker2003graph}.

\paragraph{Label Propagation}\cite{raghavan2007near}: Label Propagation identifies clusters using label or cluster information from neighboring nodes. Initially, it assigns a unique label to each node. The community label of nodes is updated based on their neighbors labels. Each node chooses to join the community to which the maximum number of its neighbors belong, with ties broken uniformly at random. At the end, nodes having the same label are grouped together as one community.  
\paragraph{Ego-splitting}\cite{epasto2017ego}: Ego-splitting is an overlapping community detection method that operates in two key steps: local partitioning of ego-networks and global graph partitioning. In both cases, it uses an iterative label propagation method. Initially, it constructs ego-nets of each node with directly  connected neighbors. Then, it applies local partitioning to create different persona graphs for each node, where each persona represents a distinct role in which the node participates. In the second step, the global partitioning process occurs on the persona graphs transformed version of the original graph where persona nodes are linked based on shared interactions—allowing detecting overlapping communities.
\paragraph{BIGCLAM (Cluster Affiliation Model for Big Networks)}\cite{yang2013overlapping}: BIGCLAM is an overlapping community detection method that identifies densely overlapping and hierarchically nested overlapping communities. The model builds on Breiger’s idea that communities emerge from shared group affiliations, representing these with a bipartite affiliation network linking nodes to the communities they belong to. Each affiliation edge has a nonnegative weight, reflecting the extent of a node’s involvement—higher weights indicate a stronger likelihood of connection with other members. Additionally, when nodes share multiple communities, each shared affiliation independently increases the probability of a connection, with stronger ties arising from overlapping memberships.

 We briefly describe other methods. \textit{Hierarchical clustering}~\cite{hastie2009elements} represents graph in multilevel structure by identifying groups of vertices with high similarity. This approach has two categories: Agglomerative algorithms that iteratively merge clusters if their similarity is sufficiently high; Divisive algorithms iteratively split clusters by removing edges connecting vertices with low similarity. In \textit{k-Clustering}~\cite{schenker2003graph}, nodes are assigned to a predifined number of $k$ clusters. Here, nodes are embedded into 2-D space, and then by maximizing or minimizing cost function, points are separated to $k$ clusters using distance between points or points to centroid.   Deep learning techniques are also incorporated into community detection algorithms. DeepWalk~\cite{perozzi2014deepwalk} is a deep learning unsupervised feature learning technique that learns social representations of a graph’s vertices by modeling a stream of short random walks. A random walk in a graph is a sequence of steps where, at each step, the walk moves from the current vertex to one of its neighboring vertices, chosen at random. The walk uniformly samples from the neighbors of the last visited vertex until it reaches the maximum length. Then it uses the Skip-gram algorithm \cite{guthrie2006closer} to maximize the probability of observing a vertex given the neighboring vertices within the random walk. The Skip-gram model is optimized using stochastic gradient descent to learn vector representations for each vertex. An extension of Deepwalk, \textit{GEMSEC}~\cite{rozemberczki2019gemsec} combines graph representation learning with clustering, aiming to enhance both tasks. The embedding keeps nodes with similar neighborhoods embedded close to each other. At the next level, it includes a clustering cost similar to $k$-means, which calculates the distance from nodes to their cluster centers. In the embedding space, it minimizes the distance to the nearest cluster center. 

\section{Proposed Framework}
\label{sec:framework}
\begin{figure}[!htbp]
    \centering
    \includegraphics[width=0.6\linewidth]{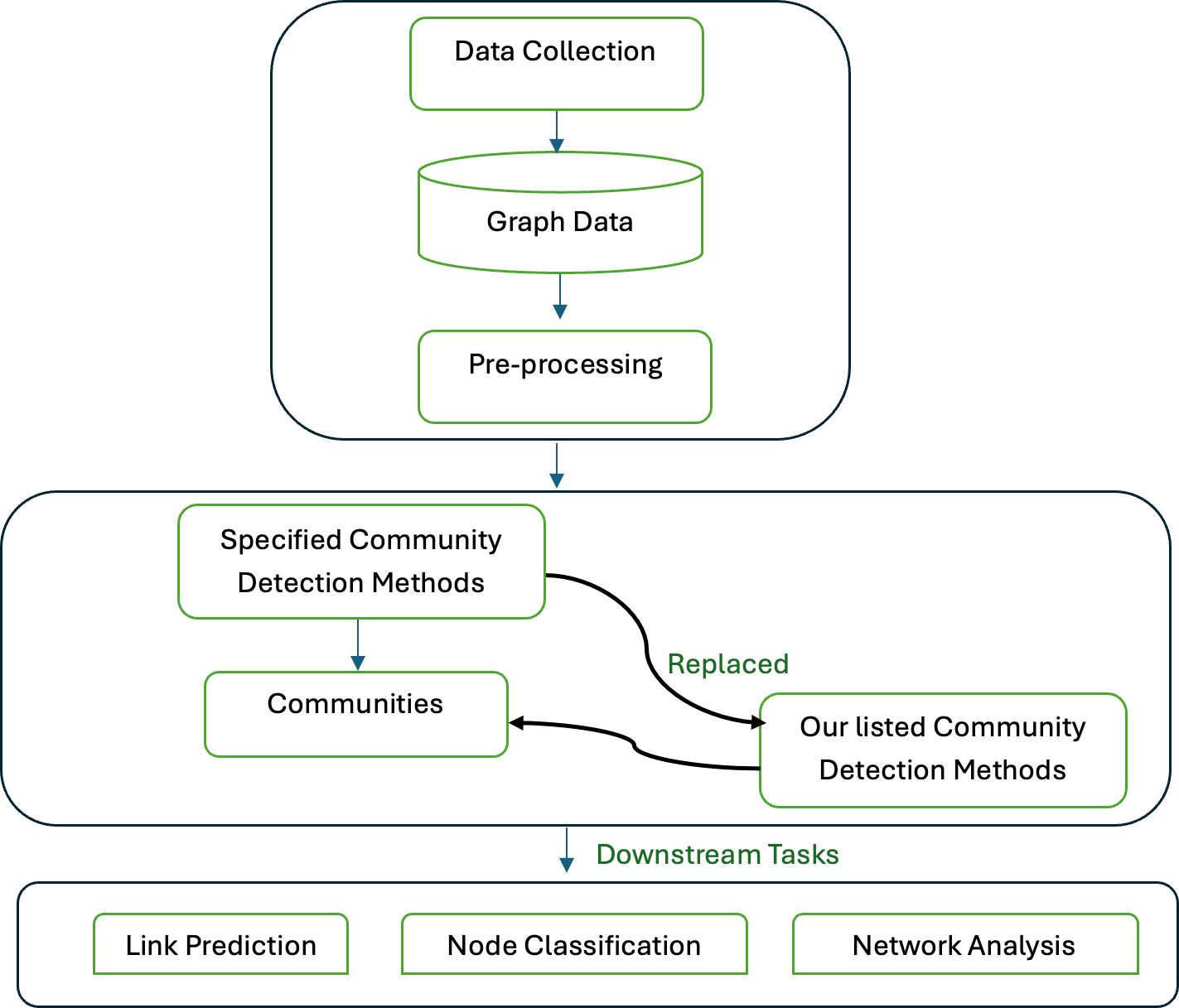}
    \caption{Overview of our Methodology}
    \label{fig:enter-label}
\end{figure}

In order to understand the impact of the chosen community detection algorithm on the performance of downstream in graph analysis and mining applications, we propose the following methodology, illustrated in Figure~\ref{fig:enter-label}. We reproduce the studies of selected published analyses and mining application which used community detection methods as an intermediate step. In these applications, we replace the community detection method proposed with a standard set of selected methods and followed the rest of the method accordingly. We will then study the performance achieved by their application on metrics appropriate for these tasks while varying the algorithm used for community detection. 

The applications perform tasks such as recommendation, node classification, or link prediction. And they usually function in a similar way. After preprocessing the data, the application performs community detection on a graph. The application then compute features that depend on the communities. These features are often based on propensities of the community as measured by a standard graph property, such as betweenness centrality of the nodes in the community. Finally, the method perform their tasks, either by making a prediction according to a simple formula or by training a machine learning model.

In order to perform the study, we needed to be able to run the described application. However, in all cases, the source code and dataset details were not documented in a way that was directly reproducible, if they were available at all. We reimplemented the  applications from scratch as faithfully as we could based on the provided description. When we differed from the described method, we noted these differences. For dataset, we sourced dataset as closely related to the task as we could, often obtaining a dataset from the same source. All the codes are available in a Git repository\footnote{\url{https://github.com/srbnghosh99/Community-Detection-in-Large-Graphs-and-Applications/tree/main}} and the associated documentation contains link to the precise dataset used in the experiments.

\section{Use-Case 1: Recommendation System}
\label{sec:usecase:recommendation}

For product recommendation system we replicated the work by Liang et al.~\cite{liang2019community}. They predict the users rating on products by using the link structure of social network. They extract features from the social network by extracting community information to account for the users' community influence. The ConSVD framework uses the community influence of users for product recommendation. 

In ConSVD framework, the first stage extracts overlapping communities using a community detection, BIGCLAM. In the next stage, it computes a measure of influence of each user $u$ within a community $c$ called propensity $\alpha_{uc}$. The algorithm takes as $\alpha_{uc}$ a centrality of $u$ within the community subgraph $c$, which could be any centrality metric such as  betweenness, closeness, and degree centrality. According to conformity theory~\cite{cialdini2004social}, users adjust their choices or preferences to align with others in the community, fostering a sense of belonging. 

The method assumes that each item $i$ can be represented by a latent $d$ dimensional vector $q_i$. The rating pattern of a community is assumed to be represented by a $d$-dimensional latent vector $p_c \in \mathbb{R}^d$. Each user also has a $d$-dimensional latent preference vector $p_u \in \mathbb{R}^d$. Liang et al. assumes a user's preference is influenced by the preference of each community proportionally to its propensity, so that the users' expressed preference is $p_u + \sum_{c:u \in c}\alpha_{uc} p_c$. 

The framework ConSVD model assumes that each user $u$ has a scalar bias $b_u$; each item $i$ has a scalar bias $b_i$; and the model has a scalar bias $\mu$. Then it models the rating of user $u$ of item $i$ by incorporating user preference and community influence:

$$ \hat{r}_{ui} = \mu + b_u + b_i + q_i^T(p_u + \sum_{c:u \in c} \alpha_{uc}p_c) $$

The model is trained by optimizing a regularized objective function.

\subsection{Dataset \& Experimental Settings}

\begin{table}[tbp]
\centering
\renewcommand\arraystretch{1.5}
\caption{Statistics of the Data Sets for Recommendation System}
\label{tab:recommendationdata}
\scriptsize
\begin{tabular}{|l|c|c|} \hline
\toprule
\rowcolor{gray!30}
Features       & {Ciao} & {Epinion}   \\ \hline\hline
\midrule
No of Users    & 7,375 & 15,396    \\ \hline
No of Items    & 105,114 &  296,277 \\ \hline
No of Ratings  & 284,086 & 922,267  \\ \hline
Trust Relations & 111,780 & 355,751  \\ \hline
\bottomrule
\end{tabular}
\end{table}

We collected data set Ciao and Epinion from Trust\footnote{\url{https://www.cse.msu.edu/~tangjili/datasetcode/truststudy.htm}}. Each user trusts a number of other users in the system. And also each has rated a number of products. Items are rated on a star scale from 1 star to 5 stars.
Table~\ref{tab:recommendationdata} shows basic statistics on the datasets. 

We tested five community detection algorithms on both datasets, including both overlapping and non-overlapping ones, and conducted the experiments as described. We conducted five-fold validations and report average performance values across the five folds. We computed the Root-Mean-Squared-Error~(RMSE) and Mean-Absolute-Error~(MAE) as performance values with lower values indicating more accurate predictions.

\begin{table}[tbp]
\centering
\begin{minipage}{0.48\textwidth}
\adjustbox{max width=\textwidth}{
\begin{tabular}{|llrr|}
\hline
\multicolumn{1}{|l|}{C.D. algorithm} & \multicolumn{1}{l|}{Centrality} & \multicolumn{1}{r|}{RMSE} & MAE \\ \hline \hline
\rowcolor{gray!30}
\multicolumn{4}{|c|}{Non-Overlapping}                                                                                                                    \\ \hline
\multicolumn{1}{|l|}{Louvain}          & \multicolumn{1}{l|}{Degree } & \multicolumn{1}{r|}{2.22}  & 1.67                     \\ \cline{2-4} 
\multicolumn{1}{|l|}{}                                   & \multicolumn{1}{l|}{Betweenness}       & \multicolumn{1}{r|}{2.46}  & 1.95                     \\ \cline{2-4} 
\multicolumn{1}{|l|}{}                                   & \multicolumn{1}{l|}{Closeness}         & \multicolumn{1}{r|}{1.89}  & 1.42                     \\ \hline
\multicolumn{1}{|l|}{Spectral}          & \multicolumn{1}{l|}{Degree } & \multicolumn{1}{r|}{2.07} & 1.52                     \\ \cline{2-4} 
\multicolumn{1}{|l|}{}                                   & \multicolumn{1}{l|}{Betweenness}       & \multicolumn{1}{r|}{2.02}  & 1.49                     \\ \cline{2-4} 
\multicolumn{1}{|l|}{}                                   & \multicolumn{1}{l|}{Closeness}         & \multicolumn{1}{r|}{1.42}  & 1.07                     \\ \hline
\multicolumn{1}{|l|}{Label Propagation} & \multicolumn{1}{l|}{Degree } & \multicolumn{1}{r|}{2.13}  & 1.56                     \\ \cline{2-4} 
\multicolumn{1}{|l|}{}                                   & \multicolumn{1}{l|}{Betweenness}       & \multicolumn{1}{r|}{2.06}  & 1.50                      \\ \cline{2-4} 
\multicolumn{1}{|l|}{}                                   & \multicolumn{1}{l|}{Closeness}         & \multicolumn{1}{r|}{\textbf{1.34}}  & \textbf{0.99}                     \\ \hline \hline
\rowcolor{gray!30}
\multicolumn{4}{|c|}{Overlapping}                                                                                                                         \\ \hline

\multicolumn{1}{|l|}{Ego-Splitting}     & \multicolumn{1}{l|}{Degree } & \multicolumn{1}{r|}{2.23}  & 1.68                     \\ \cline{2-4} 
\multicolumn{1}{|l|}{}                                   & \multicolumn{1}{l|}{Betweenness}       & \multicolumn{1}{r|}{2.26}  & 1.71                     \\ \cline{2-4} 
\multicolumn{1}{|l|}{}                                   & \multicolumn{1}{l|}{Closeness}         & \multicolumn{1}{r|}{{1.78}}  & 1.32                     \\ \hline
\multicolumn{1}{|l|}{BIGCLAM}     & \multicolumn{1}{l|}{Degree } & \multicolumn{1}{r|}{1.86}  & 1.37                     \\ \cline{2-4} 
\multicolumn{1}{|l|}{}                                   & \multicolumn{1}{l|}{Betweenness}       & \multicolumn{1}{r|}{1.98}  & 1.46                     \\ \cline{2-4} 
\multicolumn{1}{|l|}{}                                   & \multicolumn{1}{l|}{Closeness}         & \multicolumn{1}{r|}{\textbf{1.65}}  & \textbf{1.28}                     \\ \hline
\end{tabular}
}

\end{minipage}
\hfill
\begin{minipage}{0.48\textwidth}
\centering
\adjustbox{max width=\textwidth}{
\begin{tabular}{|llrr|}
\hline
\multicolumn{1}{|l|}{C.D. algorithm} & \multicolumn{1}{l|}{Centrality} & \multicolumn{1}{r|}{RMSE} & MAE \\ \hline \hline
\rowcolor{gray!30}
\multicolumn{4}{|c|}{Non-Overlapping}                                                                                                                     \\ \hline
\multicolumn{1}{|l|}{Louvain}          & \multicolumn{1}{l|}{Degree } & \multicolumn{1}{r|}{2.40}  & 1.90                     \\ \cline{2-4} 
\multicolumn{1}{|l|}{}                                   & \multicolumn{1}{l|}{Betweenness}       & \multicolumn{1}{r|}{2.23}  & 1.72                     \\ \cline{2-4} 
\multicolumn{1}{|l|}{}                                   & \multicolumn{1}{l|}{Closeness}         & \multicolumn{1}{r|}{1.76}  & 1.30                     \\ \hline
\multicolumn{1}{|l|}{Spectral}          & \multicolumn{1}{l|}{Degree } & \multicolumn{1}{r|}{1.69} & 1.27                     \\ \cline{2-4} 
\multicolumn{1}{|l|}{}                                   & \multicolumn{1}{l|}{Betweenness}       & \multicolumn{1}{r|}{1.56}  & 1.19                     \\ \cline{2-4} 
\multicolumn{1}{|l|}{}                                   & \multicolumn{1}{l|}{Closeness}         & \multicolumn{1}{r|}{\textbf{1.37}}  & \textbf{1.00}                   \\ \hline
\multicolumn{1}{|l|}{Label Propagation} & \multicolumn{1}{l|}{Degree } & \multicolumn{1}{r|}{2.22}  & 1.68                     \\ \cline{2-4} 
\multicolumn{1}{|l|}{}                                   & \multicolumn{1}{l|}{Betweenness}       & \multicolumn{1}{r|}{2.12}  & 1.58                      \\ \cline{2-4} 
\multicolumn{1}{|l|}{}                                   & \multicolumn{1}{l|}{Closeness}         & \multicolumn{1}{r|}{{1.65}}  & {1.24}                     \\ \hline \hline
\rowcolor{gray!30}
\multicolumn{4}{|c|}{Overlapping}                                                                                                                         \\ \hline 

\multicolumn{1}{|l|}{Ego-Splitting}     & \multicolumn{1}{l|}{Degree } & \multicolumn{1}{r|}{1.97}  & 1.48                     \\ \cline{2-4} 
\multicolumn{1}{|l|}{}                                   & \multicolumn{1}{l|}{Betweenness}       & \multicolumn{1}{r|}{2.07}  & 1.56                     \\ \cline{2-4} 
\multicolumn{1}{|l|}{}                                   & \multicolumn{1}{l|}{Closeness}         & \multicolumn{1}{r|}{{1.81}}  & 1.37                     \\ \hline
\multicolumn{1}{|l|}{BIGCLAM}     & \multicolumn{1}{l|}{Degree } & \multicolumn{1}{r|}{2.22}  & 1.68                     \\ \cline{2-4} 
\multicolumn{1}{|l|}{}                                   & \multicolumn{1}{l|}{Betweenness}       & \multicolumn{1}{r|}{2.12}  & 1.58                     \\ \cline{2-4} 
\multicolumn{1}{|l|}{}                                   & \multicolumn{1}{l|}{Closeness}         & \multicolumn{1}{r|}{\textbf{1.65}}  & 1.24                     \\ \hline
\end{tabular}
}
\end{minipage}
\caption{Evaluation on Product Recommendation Performance Ciao (left) Epinion (right) using different Community Detection algorithms and Centralities.}
\label{tab:reco}
\end{table}

\subsection{Results}
Table~\ref{tab:reco} presents the quality of user product rating prediction for the Ciao and Epinion datasets. The results indicate that the closeness propensity consistently yields better outcomes compared to the other two propensities. For all community detection methods, the RMSE obtained with closeness is from $0.2$ stars better to $0.8$ star better than the second best propensity.

The choice of community detection methods clearly influences the outcome of this task. For instance, in Ciao dataset, the Louvain method using Degree centrality yields values of 2.22 and 1.67, whereas the BIGCLAM method gives 1.86 and 1.37. Additionally, BIGCLAM provides competitive results for all propensity types, suggesting that overlapping features here enrich feature extraction. On the other hand, in Epinion dataset, the Spectral method gives competitive results for all propensity types, indicating that the non-overlapping features yield better characteristics. The outcomes are not only influenced by the choice of community detection algorithm but also depend on which method is more applicable for a particular dataset.

\section{Use Case 2: Trust Prediction}
\label{sec:usecase:trust}

For trust prediction, we replicated the work by Beigi et al.~\cite{beigi2014leveraging}. The paper introduced a two-phased approach to predict trust values between each pair of users using rating information given to products.

For two users $i$ and $j$, one can use the rating of products by these users $r_{ik}$ and $r_{jk}$ in order to compute a similarity $R_{ij}$ between their ratings. The similarity, is computed by $R_{ij} = \frac{\sum_{k}{r_{ik}}{r_{jk}}}
{{\sqrt{\sum_k r^2_{ik}} {\sqrt{\sum_k r^2_{jk}}}}}$. An issue with such a calculation is that for two users there may not be a single product that they have both rated, making this similarity calculation inherently inaccurate.

To remedy this calculation, Beigi et al. leverages community information in a social network. They used two community detection algorithms: Game theoretic approach and Smart Local Moving. Then, for each communities $C_1$ they name a center $ce_{C_1}$ by naming the vertex in that community with the highest value for a centrality. Beigi et al. used (a) Betweenness (b) Eigenvector (c) MaxDegree (d) MaxTrustor (indegree) (e) MaxTrustee (outdegree) (f) Random. 
The center of the communities can then be used as proxy for the behavior of users in that community. 

Beigi et al. compute the final trust prediction $P_{ij}$ between users $i$ and $j$ by leveraging for each community $C_1$ and $C_2$ that $i$ and $j$ belong to their similarity to their centers $ce_{C_1}$ and $ce_{C_2}$ and the similarity between the centers.
\begin{equation}  
\label{trust_pred}
{
P_{ij} = \max_{C_1 \in C(i), C_2 \in C(j)}
          Avg (R_{i ce_{C_1}}, R_{j ce_{C_2}}, R_{ce_{C_1} ce_{C_2}})
}
\end{equation}

\subsection{Dataset \& Experimental Settings}

\begin{table}[tbp]
\centering
{\renewcommand\arraystretch{1.5}
\caption{Statistics of the Data Sets for Trust Prediction}
\label{tab:trustnet}
\scriptsize
\begin{tabular}{|l|c|c|l|l||} \hline
\rowcolor{gray!30}
Features & Ciao & Epinions  \\ \hline\hline
No of Users                 & 6,650                         & 15,396                       \\ \hline
No of Items                 & 37,590                        & 22,160                       \\ \hline
No of Ratings               & 216,562                       & 922,263                      \\ \hline
Trust Relations              & 84,544                       & 284,590                      \\ \hline

\end{tabular}}
\end{table}

We filtered the Trust data presented in Table~\ref{tab:recommendationdata} by mapping the common users in both trustnet social network and rating dataset. Table~\ref{tab:trustnet} shows pre-processed filtered dataset.

We replaced the Smart Local Moving method  used by Beigi et al. with our five methods and followed the same prediction method otherwise. 
For both Ciao and Epinion, we randomly sampled 67K known trust relations and added 33K pairs where there was no edge in the dataset. We picked threshold for deciding if a pair trust each other of $0.45$, $0.6$, $0.6$, $0.45$, and $0.55$ for Louvain, Label Propagation, Spectral, Ego-Splitting, and BigClam respectively. For Epinion, we picked a threshold value of $0.55$ for all methods. We use standard evaluation, metrics precision, recall, F1-score and AUC. A higher F1 score indicates that the model is effective in minimizing false positives and false negatives.

\begin{table}[tbp]
\centering
\begin{minipage}{0.44\textwidth}{
\renewcommand\arraystretch{0.85} \footnotesize
\adjustbox{max width=\textwidth}{
\begin{tabular}{|c|c|c|c|c|} \hline
Propensity & Precision & Recall & F1 & AUC \\ \hline\hline
\rowcolor{gray!30}
\multicolumn{5}{|c|}{Non-overlapping} \\ \hline
\multicolumn{5}{|c|}{Louvain} \\ \hline
Betweenness & 0.7997 & 0.8184 & 0.8089 & 0.5974 \\ \hline
MaxDegree   & 0.8021 & 0.8171 & 0.8095 & \textbf{0.6020} \\ \hline
MaxTrustor  & 0.7854 & 0.8515 & 0.8171 & 0.5721 \\ \hline
MaxTrustee  & 0.7942 & 0.8254 & 0.8095 & 0.5875 \\ \hline
Random      & 0.7862 & 0.8540 & 0.8187 & 0.5739 \\ \hline
\multicolumn{5}{|c|}{Spectral} \\ \hline
Betweenness & 0.7434 & 0.5344 & 0.6218 & 0.4867 \\ \hline
MaxDegree   & 0.7526 & 0.5756 & 0.6523 & 0.5001 \\ \hline
MaxTrustor  & 0.7465 & 0.5837 & 0.6552 & 0.4904  \\ \hline
MaxTrustee  & 0.7498 & 0.5632 & 0.6432 & 0.4958 \\ \hline
Random      & 0.7512 & 0.5867 & 0.6588 & 0.4979   \\ \hline
\multicolumn{5}{|c|}{Label Propagation} \\ \hline
Betweenness & 0.7192 & 0.6752 & 0.6965 & 0.4367  \\ \hline
MaxDegree   & 0.7192 & 0.6752 & 0.6965 & 0.4367 \\ \hline
MaxTrustor  & 0.7330 & 0.7115 & 0.7221 & 0.4618  \\ \hline
MaxTrustee  & 0.7295 & 0.7024 & 0.7157 & 0.4552 \\ \hline
Random      & 0.7349 & 0.7359 & 0.7354 & 0.4643   \\ \hline \hline
\rowcolor{gray!30}
\multicolumn{5}{|c|}{Overlapping} \\ \hline
\multicolumn{5}{|c|}{Ego Splitting} \\ \hline
Betweenness & 0.7541 & \textbf{0.9994} & \textbf{0.8596} & 0.5042  \\ \hline
MaxDegree   & 0.7541 & \textbf{0.9994} & \textbf{0.8596} & 0.5042 \\ \hline
MaxTrustor  & 0.7535 & 0.9972 & 0.8584 & 0.5025  \\ \hline
MaxTrustee  & 0.7508 & 0.9930 & 0.8571 & 0.5038 \\ \hline
Random      & 0.7543 & 0.9942 & 0.8578 & 0.5048   \\ \hline
\multicolumn{5}{|c|}{BIGCLAM} \\ \hline
Betweenness & 0.70 & 0.82 & 0.75 & 0.5467  \\ \hline
MaxDegree   & 0.67 & 0.93 & 0.78 & 0.4964   \\ \hline
MaxTrustor  & 0.67 & 0.94 & 0.78 & 0.4954  \\ \hline
MaxTrustee  & 0.67 & 0.92 & 0.77 & 0.4949  \\ \hline
Random      & 0.67 & 0.93 & 0.78 & 0.5001      \\ \hline

\end{tabular}}}
\end{minipage}
\begin{minipage}{0.44\textwidth}
{\renewcommand\arraystretch{0.83} \footnotesize
\adjustbox{max width=\textwidth}{
\begin{tabular}{|c|c|c|c|c|} \hline
Propensity & Precision & Recall & F1 & AUC \\ \hline\hline
\rowcolor{gray!30}
\multicolumn{5}{|c|}{Non-overlapping} \\ \hline
\multicolumn{5}{|c|}{Louvain } \\ \hline
Betweenness & 0.6841 & 0.9370 & 0.7908 & 0.5362 \\ \hline
MaxDegree   &  0.6834 & 0.9432 & 0.7926 & 0.5351 \\ \hline
MaxTrustor  & \textbf{0.6968} & 0.9009 & 0.7858 & \textbf{0.5588} \\ \hline
MaxTrustee  & 0.6949 & 0.9038 & 0.7857 & 0.5556 \\ \hline
Random      & 0.6953 & 0.8992 & 0.7842 & 0.5560 \\ \hline
\multicolumn{5}{|c|}{Spectral} \\ \hline
Betweenness & 0.6789 & 0.6811 & 0.6800 & 0.5187 \\ \hline
MaxDegree   &  0.6792 & 0.6544 & 0.6666 & 0.5184 \\ \hline
MaxTrustor  & 0.6764 & 0.6904 & 0.6833 & 0.5153 \\ \hline
MaxTrustee  & 0.6775 & 0.6825 & 0.6800 & 0.5167 \\ \hline
Random      & 0.6741 & 0.6994 & 0.6865 & 0.5119 \\ \hline
\multicolumn{5}{|c|}{Label Propagation} \\ \hline
Betweenness & 0.6693 & 0.9852 & 0.7971 & 0.5064  \\ \hline
MaxDegree   & 0.6690 & \textbf{0.9863} & 0.7972 & 0.5057 \\ \hline
MaxTrustor  & 0.6695 & 0.9725 & 0.7931 & 0.5068  \\ \hline
MaxTrustee  & 0.6695 & 0.9852 & 0.7972 & 0.5068 \\ \hline
Random      & 0.6695 & 0.9756 & 0.7941 & 0.5067   \\ \hline \hline
\rowcolor{gray!30}
\multicolumn{5}{|c|}{Overlapping} \\ \hline
\multicolumn{5}{|c|}{Ego Splitting} \\ \hline
Betweenness & 0.6739 & {0.9772} & 0.7977 & 0.5147 \\ \hline
MaxDegree   &  0.6744 & 0.9765 & 0.7978 & 0.5160 \\ \hline
MaxTrustor  & 0.6570 & 0.8445 & 0.7390 & 0.4819 \\ \hline
MaxTrustee  & 0.6739 & 0.9750 & 0.7969 & 0.5148 \\ \hline
Random      & 0.6567 & 0.8659 & 0.7470 & 0.4809 \\ \hline
\multicolumn{5}{|c|}{BIGCLAM} \\ \hline
Betweenness & 0.6598 & 0.7334 & 0.6947 & 0.4891 \\ \hline
MaxDegree   &  0.6601 & 0.7383 & 0.6970 & 0.4895 \\ \hline
MaxTrustor  & 0.6740 & 0.9677 & 0.7945 & 0.5148 \\ \hline
MaxTrustee  & 0.6565 & 0.8109 & 0.7256 & 0.4816 \\ \hline
Random      & 0.6741 & 0.9656 & 0.7939 & 0.5150 \\ \hline
\end{tabular}}}
\end{minipage}
\caption{Evaluation on Trust Prediction on Ciao (left) and Epinion (right)}
\label{tab:trustnet_result}
\end{table}

\subsection{Results}
Tables~\ref{tab:trustnet_result} show the results of predicting trust between users using rating information from the Ciao and Epinion datasets. 

From the results, we can see that the propensity values do not have a significant impact on the outcome.  The difference between the highest F1-score or Area Under the Curve score rarely sees a fluctuation of more than $0.02$ depending on the change of propensity.

However, the community detection method has a more significant impact. In the Ciao data, there is a spread of $0.2$ in the F1 score between the spectral and Ego splitting methods. The Louvain method yields a better AUC in both datasets ($0.6020$ and $0.5588$ in Ciao and Epinion, respectively) indicating that it differentiates between classes of trust and no-trust relations more effectively than the other methods. The Ego-splitting method has better recall for the Ciao dataset, while Label Propagation performs better in the Epinion dataset. From this we can clearly state that, in this application, the chosen community detection method strongly impacts the performance of the method.

\section{Use Case 3: Anomaly Detection}
\label{sec:usecase:anomaly}

For anomaly detection, we followed the steps introduced by Keyvanpour et al.~\cite{keyvanpour2020ad}, which utilize the social network graph and its properties to detect anomalous nodes. Keyvanpour et al. applied the Louvain \cite{blondel2008fast}, Label Propagation \cite{zhur2002learning}, and Clique percolation \cite{palla2005uncovering} community detection algorithms. In this case, both weighted and unweighted graphs can be employed. 

After detecting communities using non-overlapping algorithms, auxiliary communities are identified for finding overlapping communities: Every node on the border of its community is assigned to a new community called the auxiliary community if it has a neighbor that does not belong to the same community. In the final step, the model filters nodes based on six features of the communities: (a) Number of communities, (b) Number of communities over number of neighbors, (c) 1 - cluster coefficient, (d) Deg(V)/edge weights, (e) Cliqueness, (f) Starkness. 

\textbf{Anomaly score:}
These features are computed for each node based on the communities they belong to and the graph structure. Each feature needs to be weighted based on the importance of the features to find the best combination. For each feature, there is a threshold value defined. For feature values greater than threshold, the indicator is set to 1 and for smaller values the indicator is set to zero. 
The indicator function: 
\begin{equation} 
\centering
\label{eq2}
{
    I(\phi_i(v),T_i) = 
        \begin{cases}
        1,          \phi(v) > T_i \\
        0,          otherwise
        \end{cases}
    }
\end{equation}

Here, $\phi_i(v)$ is the value of $i$th feature of node $v$.
To compute the score of anomaly, Keyvanpour et al. sum up all the features multiplied by their estimated weights using Equation~\ref{eq2}. The sum represents the prediction score of the anomaly.
\begin{equation} 
\centering
\label{eq2}
    S_c(v) = \sum w_i I(\phi_i(v),t_i)
\end{equation}

\begin{table}[tbp]
\centering
{\renewcommand\arraystretch{1.5}
\caption{Statistical information of Data Sets for Anomaly Detection}
\label{tab:anomalydata}
\scriptsize
\begin{tabular}{|l|r|r|} \hline
\rowcolor{gray!30}
{Features} & {DGraph-Fin} & {YelpHotel} \\ \hline\hline
No of Nodes & \num{3700550} & \num{2288}  \\ \hline
No of Edges & \num{3997260} & \num{41631}  \\ \hline
Number of Anomalies & \num{15509} & \num{250} \\ \hline
\end{tabular}}
\end{table}

\subsection{Dataset \& Experiments}
For detecting anomalies, we utilized the YelpHotel (HTL)~\cite{wang2023cross} dataset referenced in the previous paper. However, we appear to have collected the data at a different time since the size of the datasets are different. The YelpHotel dataset comprises online review graphs from Yelp, focused on accommodation and dining establishments in the Chicago area. Each node represents a reviewer, while an edge signifies that two reviewers have commented on the same business. If a node is marked as suspicious, it generally indicates that the associated reviewer has been identified for potentially submitting fake reviews. The YelpHotel dataset contains two categories: normal nodes and anomaly nodes.

We have also used a financial graph for anomaly detection. The DGraph~\cite{huang2022dgraph}, provided by Finvolution Group, an industry leader in China's online consumer finance with over 140 million registered consumers, assesses loan requests and determines whether to approve them. Fraudulent users are those who borrowed money but did not repay it, ignoring the platform's reminders. The nodes of Class 1 represent fraudulent users, while nodes of Class 0 represent normal users. The other two classes, 2 and 3, consist of background users. The dataset is significantly imbalanced; the distribution of labels for each class is as follows: class 0: 1210092, class 1: 15509, class 2: 1620851, class 3: 854098.
Table~\ref{tab:anomalydata} shows the dataset information.

\paragraph{Method} As in the previous experiment, we replaced the community detection algorithm used by each of our five methods. Additionally, we also consider the case where a single community contains the entire graph in order to understand the impacts of the methodology.

\paragraph{Parameter Setting}
Determining the weight value of each feature manually is time-consuming and often inaccurate. Furthermore, the linear Equation~\ref{eq2} used for calculating scores does not yield high accuracy in predictions. Therefore, instead of relying on the linear equation, we employed XGBoost~\cite{chen2016xgboost}, a machine learning model that provides a more sophisticated approach to learning the relationship between features and target variables. For binary classification problems, XGBoost achieves higher accuracy, even in the presence of class imbalance, by weighting the positive class more heavily during training, thereby making the model more sensitive to the minority class. 
We followed the same steps as before, adjusting the XGBoost model parameters for binary and multiclass problems.

\paragraph{Performance reporting} For each class, we present precision, recall, F1 score, AUC, and average accuracy. In the case of multiclass classification, the AUC is set to the One-vs-Rest~(OvR) approach, which creates a binary classifier that distinguishes the target class from all other classes.

\subsection{Results}

\paragraph{DGraph dataset.} 
Table~\ref{tab:anomalydata_dgraph} presents results on the DGraph dataset. We were unable to compute the Spectral and BIGCLAM partitioning  due to computational constraints associated with processing such a large graph. 

The results highlight a multiclass issue where the quantity of labeled anomaly nodes is considerably limited. The performance of the remaining methods indicates that the precision values are quite low, usually under 50\% and typically under 1\% for the anomalous class, Class 1. The results demonstrate that the framework is not effective for this multiclass issue. No improvement was observed across the different community detection methods, suggesting that community detection approaches do not influence classification. We conjecture that this ineffectiveness stems from the graph being extremely sparse. Thus, we can conclude that this method is unsuitable for addressing the anomaly detection challenge presented by that dataset. 
\begin{table}[tbp]
\centering
{\renewcommand{\arraystretch}{1.1}
\caption{DGraph Anomaly detection average accuracy on 5-fold validation}
\label{tab:anomalydata_dgraph}
\begin{adjustbox}{width=.8\textwidth}
\begin{tabular}{|l|c|c|c|c|c|c|c|}
\hline
Method & Class & Precision & Recall & F1 & Support & AUC & Accuracy \\ \hline\hline

\multirow{4}{*}{Single Community}
  & Class 0 & 0.450983 & 0.525047 & 0.485205 & 1210092 & \multirow{4}{*}{0.64} & \multirow{4}{*}{33.49\%} \\ \cline{2-6}
  & Class 1 & 0.005130 & 0.276998 & 0.010069 & 15509   &                      &                          \\ \cline{2-6}
  & Class 2 & 0.531645 & 0.282660 & 0.369087 & 1620851 &                      &                          \\ \cline{2-6}
  & Class 3 & 0.239029 & 0.165952 & 0.194187 & 854098  &                      &                          \\ \hline

\rowcolor{gray!30}\multicolumn{8}{|c|}{Non-Overlapping Method} \\ \hline

\multirow{4}{*}{Louvain}
  & Class 0 & 0.476579 & 0.441382 & 0.458306 & 1210092 & \multirow{4}{*}{0.64} & \multirow{4}{*}{31.71\%} \\ \cline{2-6}
  & Class 1 & 0.004515 & 0.349741 & 0.008909 & 15509   &                      &                          \\ \cline{2-6}
  & Class 2 & 0.517708 & 0.335346 & 0.346595 & 1620851 &                      &                          \\ \cline{2-6}
  & Class 3 & 0.258498 & 0.105968 & 0.101218 & 854098  &                      &                          \\ \hline

\multirow{4}{*}{Label Propagation}
  & Class 0 & 0.465875 & 0.480381 & 0.468352 & 1210092 & \multirow{4}{*}{0.64} & \multirow{4}{*}{35.22\%} \\ \cline{2-6}
  & Class 1 & 0.004497 & 0.291273 & 0.008674 & 15509   &                      &                          \\ \cline{2-6}
  & Class 2 & 0.529325 & 0.421311 & 0.441418 & 1620851 &                      &                          \\ \cline{2-6}
  & Class 3 & 0.261714 & 0.040891 & 0.063994 & 854098  &                      &                          \\ \hline

\rowcolor{gray!30}\multicolumn{8}{|c|}{Overlapping Method} \\ \hline

\multirow{4}{*}{Ego Splitting}
  & Class 0 & 0.476469 & 0.441344 & 0.458234 & 1210092 & \multirow{4}{*}{0.64} & \multirow{4}{*}{38.36\%} \\ \cline{2-6}
  & Class 1 & 0.004829 & 0.267654 & 0.009409 & 15509   &                      &                          \\ \cline{2-6}
  & Class 2 & 0.539777 & 0.527036 & 0.528451 & 1620851 &                      &                          \\ \cline{2-6}
  & Class 3 & 0.298053 & 0.031912 & 0.057622 & 854098  &                      &                          \\ \hline

\end{tabular}
\end{adjustbox}
}
\end{table}

\paragraph{YelpHotel Dataset.}
Table~\ref {tab:anomalydata_yelp} shows that grouping all nodes in a single community has an accuracy of 90\%. It classifies normal nodes reasonnably well, (Precision 95\%, Recall 93\%), it struggles more recognizing anomalous nodes (Precision: 53\%, Recall 62\%).

After applying community detection methods, Louvain and Spectral perform better than the overlapping methods in terms of AUC, precision, and recall for anomaly detection. And they all perform better than using a single community. The overlapping methods (Ego Splitting and BIGCLAM) perform well in terms of recall for anomaly nodes, but they suffer from lower precision and AUC, indicating that while they identify more anomalies, they also produce more false positives.

Overall, in detecting anomalies on the YelpHotel dataset, we have shown that community detection algorithm help with the task. Moreover, the community detection algorithm chosen significantly improve the performance of the anomaly detection method.

\begin{table}[tbp]
\centering
{\renewcommand{\arraystretch}{1.1}
\caption{YelpHotel Anomaly detection average accuracy on 5-fold validation}
\label{tab:anomalydata_yelp}
\begin{adjustbox}{width=.8\textwidth}
\begin{tabular}{|l|c|c|c|c|c|c|c|}
\hline
Method & Class & Precision & Recall & F1 & Support & Accuracy & AUC \\ \hline\hline

\multirow{2}{*}{Single Community} & Normal  & 0.95 & 0.93 & 0.94 & 2035 & \multirow{2}{*}{90\%} & \multirow{2}{*}{0.62} \\ \cline{2-6}
 & Anomaly  & 0.53 & 0.62 & 0.57 & 250 & & \\ \hline\hline

\rowcolor{gray!30}
\multicolumn{8}{|c|}{Non-Overlapping Method} \\ \hline

\multirow{2}{*}{Louvain} & Normal  & 0.97 & 0.95 & 0.96 & 2035 & \multirow{2}{*}{93\%} & \multirow{2}{*}{0.71} \\ \cline{2-6}
 & Anomaly  & 0.65 & 0.74 & 0.69 & 250 & & \\ \hline

\multirow{2}{*}{Spectral} & Normal  & 0.97 & 0.96 & 0.96 & 2035 & \multirow{2}{*}{94\%} & \multirow{2}{*}{0.63} \\ \cline{2-6}
 & Anomaly  & 0.69 & 0.75 & 0.72 & 250 & & \\ \hline

\multirow{2}{*}{Label Propagation} & Normal  & 0.96 & 0.95 & 0.96 & 2035 & \multirow{2}{*}{92\%} & \multirow{2}{*}{0.64} \\ \cline{2-6}
 & Anomaly  & 0.64 & 0.72 & 0.67 & 250 & & \\ \hline\hline

\rowcolor{gray!30}
\multicolumn{8}{|c|}{Overlapping Method} \\ \hline

\multirow{2}{*}{Ego Splitting} & Normal  & 0.96 & 0.93 & 0.95 & 2035 & \multirow{2}{*}{91\%} & \multirow{2}{*}{0.64} \\ \cline{2-6}
 & Anomaly  & 0.56 & 0.72 & 0.62 & 250 & & \\ \hline

\multirow{2}{*}{BIGCLAM} & Normal  & 0.97 & 0.93 & 0.95 & 2035 & \multirow{2}{*}{91\%} & \multirow{2}{*}{0.57} \\ \cline{2-6}
 & Anomaly  & 0.58 & 0.73 & 0.66 & 250 & & \\ \hline

\end{tabular}
\end{adjustbox}
}
\end{table}




 


\section{Conclusion}
\label{sec:ccl}
This paper investigates the impact of community detection methods on downstream task performance. We propose a framework to compare performance on downstream tasks using overlapping and non-overlapping community detection techniques. The results demonstrate that the choice of community detection methods significantly influences task outcomes, suggesting the presence of hidden structural features that benefit downstream analysis. These findings underscore the necessity of informed and application-driven selection of community detection algorithms rather than arbitrary choices.  As part of future work, we aim to explore which hidden features contribute most effectively to specific types of task analyses.

\bibliographystyle{vancouver}

\bibliography{references_db}

\begin{thebibliography}{10}

\bibitem{su2022comprehensive}
Su X, Xue S, Liu F, Wu J, Yang J, Zhou C, et~al.
\newblock A comprehensive survey on community detection with deep learning.
\newblock IEEE TNNLS. 2022.

\bibitem{han2023link}
Han C, Fu X, Liang Y.
\newblock Link Prediction and Node Classification on Citation Network.
\newblock In: Proc. of IEEE ICSECE; 2023. p. 428-31.

\bibitem{bhagat2011node}
Bhagat S, Cormode G, Muthukrishnan S.
\newblock Node classification in social networks.
\newblock Social network data analytics. 2011:115-48.

\bibitem{zhang2018link}
Zhang M, Chen Y.
\newblock Link prediction based on graph neural networks.
\newblock Advances in neural information processing systems. 2018;31.

\bibitem{rehman2012graph}
Rehman SU, Khan AU, Fong S.
\newblock Graph mining: A survey of graph mining techniques.
\newblock In: Proc. of ICDIM; 2012. p. 88-92.

\bibitem{jin2021survey}
Jin D, Yu Z, Jiao P, Pan S, He D, Wu J, et~al.
\newblock A survey of community detection approaches: From statistical modeling
  to deep learning.
\newblock IEEE TKDE. 2021;35(2):1149-70.

\bibitem{von2007tutorial}
Von~Luxburg U.
\newblock A tutorial on spectral clustering.
\newblock Statistics and computing. 2007;17:395-416.

\bibitem{perozzi2014deepwalk}
Perozzi B, Al-Rfou R, Skiena S.
\newblock Deepwalk: Online learning of social representations.
\newblock In: Proc. of SIGKDD; 2014. p. 701-10.

\bibitem{cui2020adaptive}
Cui G, Zhou J, Yang C, Liu Z.
\newblock Adaptive Graph Encoder for Attributed Graph Embedding.
\newblock In: Proceedings of SIGKDD 2020; 2020. .

\bibitem{dhillon2004kernel}
Dhillon IS, Guan Y, Kulis B.
\newblock Kernel k-means: spectral clustering and normalized cuts.
\newblock In: Proc. of ACM SIGKDD; 2004. p. 551-6.

\bibitem{saramaki2007generalizations}
Saram{\"a}ki J, Kivel{\"a} M, Onnela JP, Kaski K, Kertesz J.
\newblock Generalizations of the clustering coefficient to weighted complex
  networks.
\newblock Physical Review E—Statistical, Nonlinear, and Soft Matter Physics.
  2007;75(2):027105.

\bibitem{chakraborty2017metrics}
Chakraborty T, Dalmia A, Mukherjee A, Ganguly N.
\newblock Metrics for community analysis: A survey.
\newblock ACM CSUR. 2017;50(4):1-37.

\bibitem{blondel2008fast}
Blondel VD, Guillaume JL, Lambiotte R, Lefebvre E.
\newblock Fast unfolding of communities in large networks.
\newblock Journal of statistical mechanics: theory and experiment.
  2008;2008(10):P10008.

\bibitem{donath1973lower}
Donath WE, Hoffman AJ.
\newblock Lower bounds for the partitioning of graphs.
\newblock IBM Journal of Research and Development. 1973;17(5):420-5.

\bibitem{schenker2003graph}
Schenker A, Last M, Bunke H, Kandel A.
\newblock Graph representations for web document clustering.
\newblock In: Proc. of IbPRIA; 2003. p. 935-42.

\bibitem{raghavan2007near}
Raghavan UN, Albert R, Kumara S.
\newblock Near linear time algorithm to detect community structures in
  large-scale networks.
\newblock Physical Review E—Statistical, Nonlinear, and Soft Matter Physics.
  2007;76(3):036106.

\bibitem{epasto2017ego}
Epasto A, Lattanzi S, Paes~Leme R.
\newblock Ego-splitting framework: From non-overlapping to overlapping
  clusters.
\newblock In: Proc. of ACM SIGKDD; 2017. p. 145-54.

\bibitem{yang2013overlapping}
Yang J, Leskovec J.
\newblock Overlapping community detection at scale: a nonnegative matrix
  factorization approach.
\newblock In: Proc. of ACM WSDM; 2013. p. 587-96.

\bibitem{hastie2009elements}
Hastie T, Tibshirani R, Friedman JH, Friedman JH.
\newblock The elements of statistical learning: data mining, inference, and
  prediction. vol.~2.
\newblock Springer; 2009.

\bibitem{guthrie2006closer}
Guthrie D, Allison B, Liu W, Guthrie L, Wilks Y.
\newblock A closer look at skip-gram modelling.
\newblock In: LREC. vol.~6; 2006. p. 1222-5.

\bibitem{rozemberczki2019gemsec}
Rozemberczki B, Davies R, Sarkar R, Sutton C.
\newblock Gemsec: Graph embedding with self clustering.
\newblock In: Proc. of IEEE/ACM ASONAM; 2019. p. 65-72.

\bibitem{liang2019community}
Liang B, Xu B, Wu X, Wu D, Yang D, Xiao Y, et~al.
\newblock A community-based collaborative filtering method for social
  recommender systems.
\newblock In: Proc. of IEEE ICWS; 2019. p. 159-62.

\bibitem{cialdini2004social}
Cialdini RB, Goldstein NJ.
\newblock Social influence: Compliance and conformity.
\newblock Annu Rev Psychol. 2004;55(1):591-621.

\bibitem{beigi2014leveraging}
Beigi G, Jalili M, Alvari H, Sukthankar G.
\newblock Leveraging community detection for accurate trust prediction.
\newblock In: Proc. of ASE SocialCom; 2014. .

\bibitem{keyvanpour2020ad}
Keyvanpour MR, Shirzad MB, Ghaderi M.
\newblock AD-C: a new node anomaly detection based on community detection in
  social networks.
\newblock International Journal of Electronic Business. 2020;15(3):199-222.

\bibitem{zhur2002learning}
Zhu X, Ghahramani Z.
\newblock Learning from labeled and unlabeled data with label propagation.
\newblock Carnegie Mellon University; 2002. CMU-CALD-02-107.

\bibitem{palla2005uncovering}
Palla G, Der{\'e}nyi I, Farkas I, Vicsek T.
\newblock Uncovering the overlapping community structure of complex networks in
  nature and society.
\newblock Nature. 2005;435(7043):814-8.

\bibitem{wang2023cross}
Wang Q, Pang G, Salehi M, Buntine W, Leckie C.
\newblock Cross-domain graph anomaly detection via anomaly-aware contrastive
  alignment.
\newblock In: Proc. of AAAI. vol.~37; 2023. p. 4676-84.

\bibitem{huang2022dgraph}
Huang X, Yang Y, Wang Y, Wang C, Zhang Z, Xu J, et~al.
\newblock Dgraph: A large-scale financial dataset for graph anomaly detection.
\newblock Advances in Neural Information Processing Systems. 2022;35:22765-77.

\bibitem{chen2016xgboost}
Chen T, Guestrin C.
\newblock Xgboost: A scalable tree boosting system.
\newblock In: Proc. of ACM SIGKDD; 2016. p. 785-94.

\end{thebibliography}

\end{document}